------------------------------------------------------------------------
\date{}
\documentstyle{article}
\begin{document}
\Large
\begin{center}
{\bf Nonfactorizable contributions to weak $D\rightarrow PV$ decays}
\end{center}
\vskip 2truecm \begin{center} \large \baselineskip 24pt
{\bf K. K. Sharma$^{\dag}$, A. C. Katoch$^{\dag}$ and R. C. Verma$^{1}$}\\
International Centre of Theoretical Physics, Trieste, Italy.\\
$^{\dag}$Centre of Advanced Study in Physics, Department of Physics \\
Panjab University, Chandigarh-160 014, India.\\
\end{center}
\vskip 3truecm
\begin{center}
{\bf ABSTRACT}
\end{center}
\large \baselineskip 24pt \hskip 0.5truecm
We investigate nonfactorizable contributions to two-body hadronic decays
of the charmed mesons to a pseudoscalar meson and a vector meson in
Cabibbo-favored mode. Employing SU(3)-flavor symmetry for the nonfactorizable
matrix elements, we obtain branching ratios of the decays in consistent
agreement
with experiment.
\vskip 1.0cm
$^{1}$Permanent Address: Centre of Advanced Study in Physics, Department of
Physics,
Panjab University, Chandigarh-160 014, India.\\
\newpage \large
\begin{center}
{\bf I.  Introduction}
\end{center}
\par \baselineskip 24pt
With the availability of extensive data on two-body weak hadronic decays of
heavy flavor mesons [1], it has now become possible to test the
validity of the factorization model which has been considered to be
supported by the $D$ meson phenomenology [2-4]. In the recent works [5-8], it
has been shown that the factorization model fails to
account for the observed data on charmed and bottom meson hadrons. Large
$N_{c}$ limit,
in which nonfactorizable contributions are usually ignored,
does not work when extended to $B$ meson decays as these clearly demand [6] a
positive value of the QCD parameter $a_{2}$. Even in the $D$ meson decays,
universal choice of the parameters $a_{1}$ and $a_{2}$ does not explain
many of the hadronic decays of $D$ and $D_{s}$ mesons.
For instance, the measured branching ratios of $\eta$ and $\eta'$ emitting
Cabibbo-angle-favored decays of charmed mesons are considerably larger than
those predicted in the spectator quark picture.
Annihilation terms, if used to bridge the discrepancy between
theory and experiment, require large form factors particularly
for $D^{0} \rightarrow \eta / \eta' ~+~ \bar K^{0}$ and
$D^{0} \rightarrow \eta  ~ +~ \bar
K^{*0}$ decays [7]. Further, factorization also fails to relate
$D_{s}^{+} \rightarrow \eta / \eta' ~ +~
\pi^{+}/\rho^{+}$ decays  with semileptonic decays $D_{s}^{+}
\rightarrow \eta / \eta'~+~ e^{+} ~ +~ \nu $ [7, 8] in a consistent manner.
\par
Recently, there has been a growing
interest in studying the nonfactorizable terms for weak hadronic
decays of the heavy flavor hadrons. Many attempts have been made to estimate
the amount of nonfactrizable effects needed to reproduce the experimental
results
for charmed and bottom sector [9-11] when real value $N_{c} = 3$ is used. In an
earlier
work [12], using isospin symmetry one of us (RCV) has searched for a
systematics in these
 estimates for
 various decays of $D^{+}$ and $D^{0}$ mesons. It has been shown that the
nonfactorizable
 isospin $1/2$ and $3/2$ reduced ampiltudes may bear a universal ratio for
 $D\rightarrow \bar K \pi /\bar K \rho/ \bar K^{*} \pi /\bar K a_{1} /
\bar K ^{*} \rho $ decay modes. The formalism, when
generalized to $SU(3)$ for studying $D \rightarrow PP$ decays, where $P$
denotes a pseudoscalar
meson, has resulted in a consistent fit with experiment [13].
\par
In this work,  we have extended the
SU(3)-flavor analysis of nonfactorizable contributions to
$D\rightarrow PV$ decays, where V denotes a vector meson.
We analyze the
Cabibbo- favored decays of $D^{0},~ D^{+}$ and $D_{s}^{+}$ mesons
 taking into account the final state interactions (FSI). In section II, we
develop
  the formalism. Results and discussion are given in the last section.
\begin{center}
{\bf II.  Formalism} \end{center}
We start with the
effective weak Hamiltonian
$$ H_{w} ~~ = ~~
\tilde{G}_{F} [c_{1} ( \bar u d)( \bar s c) + c_{2} ( \bar s d) ( \bar u
c) ], \eqno(1)$$
 where $ \tilde{G}_{F} = \frac {G_{F}} { \sqrt2}
V_{ud} V_{cs}^{*} $ and $ \bar q_{1} q_{2} (\equiv \bar q_{1}
\gamma_{\mu} (1 - \gamma_{5} ) q_{2})$ represents color singlet V - A current
and the QCD
coefficients at the charm mass scale are
$$ c_{1} = 1.26 \pm 0.04,
\hskip 2.5 cm c_{2} = - 0.51 \pm 0.05. \eqno(2)$$
 Due to the Fierz
transformation of the product of two Dirac currents in (1) in $
N_{c}$-color space, the Hamiltonian takes the following form [9]:
$$ H_{w} ^{CF} ~~ = ~~ \tilde{G}_{F} [a_{1} (
\bar u d)( \bar s c) + c_{2} H_{w}^{8} ], $$
$$ H_{w} ^{CS} ~~ = ~~ \tilde{G}_{F} [a_{2} ( \bar s d ) ( \bar u c ) + c_{1}
\tilde{H}_{w}^{8} ], \eqno(3)$$
for color favored (CF) and color
suppressed (CS) decay amplitudes respectively. Here,
$$ a_{1,2} = c_{1,2} + \frac {c_{2,1}} {N_{c}}, \eqno(4)$$
$$ H^{8}_{w}~~=~~ \frac {1} {2} \sum_{a=1}^{8} ( \bar u \lambda ^{a} d ) (
\bar s \lambda ^{a} c ),$$
$$\tilde{H} ^ {8}_ {w} ~~= ~~ \frac {1} {2} \sum_{a=1}^{8} ( \bar s \lambda
^{a} d ) (
\bar u \lambda ^{a} c ),\eqno(5)$$
where $ \bar q_{1} \lambda ^{a} q_{2}
(\equiv \bar q _{1} \gamma_{\mu} (1 - \gamma_{5} ) \lambda ^{a}
q_{2}) $ represents color octet current.
\par
Matrix elements of the first terms in (3) can be calculated using the
factorization scheme.  These are given column (ii) of Table I. So long
as one restricts to the color-singlet intermediate states, second
terms in (3) are ignored and one usually treats $a_{1}$ and
$a_{2}$ as input parameters in place of using $N_{c} = 3 $ in
reality. It is generally believed [2-4] that the $ D \rightarrow
\bar K\pi$ decays favor $N_{c} \rightarrow \infty $ limit, i.e.,
$$a_{1} \approx 1.26, ~~~ a_{2} \approx -0.51. \eqno(6)$$\\
However, it has been shown that this does not explain all the
decay modes of charm mesons [5,7]. For instance, the observed
$D^{0}
\rightarrow \eta \bar {K^{0}} $ and $ D^{0} \rightarrow \eta' \bar K^{0}
$ decay widths are larger than those predicted in
the spectator quark model. Also in $ D \rightarrow PV $ mode,
measured branching ratios for $ D^{0} \rightarrow \bar K^{0} \omega~ /
\eta \bar K^{*0} $, $D^{+} \rightarrow \bar K^{*0} \pi^{+}$, and
 $ D_{s}^{+} \rightarrow \bar K^{0} K^{*+}~ / \eta \rho^{+}~ / \eta' +
\rho^{+},$
 are considerably higher than those predicted by
the spectator quark diagrams. In addition to the spectator quark diagram,
factorizable W-exchange or W-annihilation diagrams may contribute to the weak
nonleptonic
decays. However, such contributions are normally expected to be suppressed [2]
in
the meson decays. For $D^{+}$ meson decays, these do not appear in the
Cabibbo favored decay process. For $D^{0}$ meson
decays, these are further color-suppressed as these involve lower QCD
coefficient $a_{2}$. Therefore, we have ignored them in the present analysis.
\par
We now investigate nonfactorizable contributions to these decays.
Matrix elements of $ {H}^{8}_{w}$ and $ \tilde{H}^{8}_{w}$
between charm mesons and two-body uncharmed final states are
difficult to calculate theoretically [9,10], as these involve
nonperturbative effects arising due to soft-gluon exchange. We
employ SU(3)-flavor-symmetry [14] to handle these matrix
elements. In the SU(3) limit, the two Hamiltonians
$ {H}^{8}_{w}$ and $ \tilde {H}^{8}_{w}$ behave like $ {H}^{2}_{13}$
component of $ 6^{*}$ and 15 representations of the SU(3). Since
$ {H}^{8}_{w}$ and $ \tilde {H}^{8}_{w}$ transform into each other
under the interchange of $u $ and $ s $ quarks, which forms V-spin subgroup
of the flavor-SU(3), the reduced matrix elements satisfy
$$ <P V || \tilde {H}^{8}_w || D > ~=~<P V || H^{8}_w || D >. \eqno(7) $$\\
The matrix elements $< PV | H^{8}_{w} | D >$, appearing in the
nonfactorizable effects, are considered as $ weak~
spurion + D~meson \rightarrow P~ +~ V $ scattering process, whose
general structure can be written as
$$ < PV | H_{w}^{8} | D > ~~ = ~~
[a_{1}(P^{b}_{m}V^{m}_{a}P^{c}) +
a_{2}(P^{m}_{a}V^{b}_{m}P^{c})$$
$$\hskip 3truecm +
a_{3}(P^{b}_{a}V^{c}_{m} + P^{b}_{m}V^{c}_{a})P^{m}]H
^{a}_{[b,c]} $$
$$\hskip 3truecm+
[b_{1}(P^{b}_{m}V^{m}_{a}P^{c}) +
b_{2}(P^{m}_{a}V^{b}_{m}P^{c})$$
$$\hskip 3truecm +
b_{3}(P^{b}_{a}V^{c}_{m}P^{m})
+b_{4}(P^{b}_{m}V^{c}_{a}P^{m})]H^{a}_{(b,c)} $$
$$\hskip
3truecm+ [e_{1}(P^{c}_{a}V^{m}_{m}P^{b})
+e_{2}(P^{m}_{m}V^{c}_{a}P^{b})]H^{a}_{[b,c]}$$
$$\hskip 3truecm
+ [d_{1}(P^{c}_{a}V^{m}_{m}P^{b})
+d_{2}(P^{m}_{m}V^{c}_{a}P^{b})]H^{a}_{(b,c)}, \eqno(8) $$
 where $ P^{a} $ denotes triplet of D-mesons $ P^{a} \equiv(D^{0},~ D^{+},~
 D_{s}^{+})$ and $P_{b}^{a}$,
$V^{a}_{b}$ denote $ 3 \otimes 3 $ matrices of uncharmed
pseudoscalar meson and vector meson nonets respectively. For pseudoscalar
mesons,
 $$ P_{b}^{a} ~=~
\left(\matrix{ P^{1}_{1} &\pi^{+} &K^{+}\cr \pi^{-} & P^{2}_{2}
&K^{0}\cr K^{-} &\bar K^{0} &P^{3}_{3}\cr}\right) \eqno(8)$$
with
$$ P^{1}_{1} ~ =~ \frac {1} {\sqrt2}
(\pi^{0} + \eta \sin{\theta} + \eta' \cos{\theta}),$$
$$ P^{2}_{2}~ = ~\frac {1} {\sqrt2}
(-\pi^{0} + \eta \sin{\theta} + \eta' \cos{\theta}),$$
$$ P^{3}_{3}~ =~ -\eta \cos{\theta} + \eta'
\sin{\theta}, \eqno(9) $$
where $\theta$ governs the $\eta-\eta'$ mixing,
and is related to the physical mixing [1] as,
$$\theta ~ = ~ {\theta}_{ideal} - {\phi}_{Phy}. \eqno(10)$$\\
For vector meson nonet,
 $$ V_{b}^{a} ~~ =~~ \left(\matrix{
V^{1}_{1} &\rho^{+} &K^{*+}\cr \rho^{-} &V^{2}_{2} &K^{*0}\cr
K^{*-} &\bar K^{*0} &V^{3}_{3}\cr}\right) \eqno(11)$$
 with
 $$ V^{1}_{1} ~ = ~ \frac {1} {\sqrt2}
(\rho^{0} + \omega  \cos{\theta}' + \phi \sin{\theta}')),$$
$$ V^{2}_{2}~ =~ \frac {1} {\sqrt2}
(-\rho^{0} + \omega  \cos{\theta}' + \phi \sin{\theta}')) ,$$
$$ V^{3}_{3}~  = ~  \omega \sin{\theta}' - \phi \cos{\theta}'.$$
For ideal $\omega - \phi$ mixing
$$ \theta ' ~=~ 0.  \eqno(12) $$
\par
In addition to the nonfactorizable effects considered so far, there may
also arise nonfactorizable effects involving product of color-singlet currents.
However, these may be relatively
suppressed [10]. Even if these are included here, it has been shown [13] that
their
contributions can be absorbed in the unknown reduced amplitudes appearing in
(7)
due to the similar structure in the SU(3) framework.
\par
There exists a straight correspondence between the terms appearing in
(7) and various quark level processes. The terms
involving the coefficients $a_{1}, ~ a_{2}, ~ b_{1}~ {\rm and} ~b_{2}$
represent annihilation diagrams. Notice that, due to the involvement of gluons,
these are no longer  suppressed. The
terms having coefficients  $a_{3}, ~ b_{3}$ and
$b_{4}$ represent spectator-quark like diagrams where the
uncharm quark in the parent D-meson flows into one of the final
state mesons. The last terms having coefficients
 $e_{i}'s $ and $d_{i}'s $ are hair-pin diagrams, where
$ q \bar q $ pair generated in the process hadronizes to one of the
final state mesons. Choosing $ H_{13}^{2}$ component from
(7), we obtain nonfactorizable contributions to various $ D
\rightarrow PV$ decays. These are given in column (ii) of Table II.
\vskip 0.5 cm \begin{center} {\bf III. Results and Discussion} \end{center}
\par
The decay rate formula for $D\rightarrow PV$ decays is given by
$$\Gamma (D\rightarrow PV) ~~ = ~~
|\tilde{G}_{F}|^{2} \frac {p_{c}^{3}} {8\pi
m^2_{V}}|A(D\rightarrow PV)|^{2},\eqno(13)$$
where $p_{c}$ is the three-momentum of final state particles in the rest frame
of
D meson and $m_{V}$ is the mass of vector meson emitted. Now we proceed to
determine nonfactorizable effects to various decays. First,
 we determine the factorizable contributions to various
decays using $N_{c} = 3 $ which fixes,
 $$ a_{1} ~ = ~1.09, ~~ a_{2} ~=~ -0.09, \eqno(14)$$
ignoring the errors in the QCD coefficients $c_{1}$ and $ c_{2}$. For the form
factors at zero momentum transfer, we use
$$ F^{DK}_{1}(0) ~ = ~ 0.76,~~ F^{D\pi}_{1}(0) ~ =~0.83,$$
$$ A^{DK^{*}}_{0}(0) ~=~ 0.75, \eqno(15)$$
as guided by the semileptonic decays of $D-$ mesons [1, 15-17], and
 $$ F^{D\eta}_{1}(0) ~ = ~ 0.681, ~~
F^{D\eta'}_{1}(0) ~ =~ 0.655,$$
$$ F^{D_{s}\eta}_{1}(0)~ = ~ 0.723, ~~
 F^{D_{s}\eta'}_{1}(0) ~ = ~ 0.704,$$
$$ F^{D_{s}K}_{1}(0) ~= ~ 0.760; \eqno(16) $$
$$ A^{D\rho}_{0}(0) ~=~ 0.669, ~~
A^{D\omega}_{0}(0) ~= ~ 0.669, $$
$$ A^{D\phi}_{0}(0)~ =~0.669,~~
A^{D_{s}\omega}_{0}(0) ~ =~ 0.700,$$
$$ A^{D_{s}\phi}_{0}(0) ~=~0.700, ~~
A^{D_{s}K^{*}}_{0}(0) ~ =~ 0.634, \eqno(17)$$
from the BSW model [2] for want of experimental information. These  form
factors
are then extraploated in $ q^{2} $ using a monopole form
with appropriate pole masses. Following values are used for the meson decay
 constants [1,2] (in GeV)
 $$ f_{\pi}~ =~ 0.132,~~
 f_{K} ~ = ~0.161,$$
 $$ f_{\rho} ~ = ~ 0.212, ~~
 f_{K^{*}} ~=~ 0.221. \eqno(18) $$
Numerical values of the factorizable amplitudes are given in col (iii)
of Table I.
\par
Notice that $ D \rightarrow \bar K \rho $ and $ D
\rightarrow \bar K^{*} \pi$ decays involve elastic FSI whereas
the remaining decays are not affected by them. As a result, the
isospin amplitudes 1/2 and 3/2 appearing in these
decays may develop different phases. We illustrate the procedure for $ D
\rightarrow
 \bar K \rho$ amplitudes:
$$ A(D^{0}\rightarrow K^{-} \rho^{+} )
{}~~= ~~ \frac {1} { \sqrt3} [ A_{3/2} e^{i
\delta_{3/2}} + \sqrt2 A_{1/2} e^{i \delta_{1/2}} ], $$
$$ A(D^{0}
\rightarrow \bar K^{0} \rho^{0} ) ~~ =~~ \frac
{1} { \sqrt3} [ \sqrt2 A_{3/2} e^{i \delta_{3/2}} - A_{1/2} e^{i
\delta_{1/2}} ], $$
$$ A(D^{+} \rightarrow \bar K^{0} \rho^{+} )
{}~~ = ~~ \sqrt3 A_{3/2} e^{i \delta_{3/2}}.
\eqno(19) $$
Following phase independent relations:
$$ | A(D^{0} \rightarrow
K^{-} \rho^{+} ) |^{2} + | A ( D^{0} \rightarrow \bar K^{0}
\rho^{0} )|^{2} ~~= ~~| A_{1/2} |^{2} + | A_{3/2}
|^{2},$$
$$ | A(D^{+} \rightarrow \bar K^{0} \rho^{+} ) |^{2}
{}~~ = ~~ 3 | A_{3/2} |^{2}, \eqno(20)$$
allow us to work without the phases. Writing the
total decay amplitude as a sum of the factorizable and nonfactorizable
parts
$$ A ( D \rightarrow PV) ~=~ A^{f} ( D \rightarrow PV) ~+~
A^{nf} ( D \rightarrow PV), \eqno(21)$$
we obtain
$$A_{1/2}^{nf}(D\rightarrow \bar K \rho)~ = ~
\frac{1}{ \sqrt3} \{ \sqrt2 A^{nf} (D^{0} \rightarrow K^{-} \rho^{+} ) -
A^{nf} (D^{0} \rightarrow \bar K^{0} \rho^{0}) \}, \eqno(22)$$
$$ A_{3/2}^{nf}(D\rightarrow \bar K \rho ) ~ =~
 \frac{1}{ \sqrt3} \{ A^{nf} (D^{0} \rightarrow K^{-} \rho^{+} ) +
\sqrt2 A^{nf} (D^{0} \rightarrow \bar K^{0} \rho^{0}) \}$$ $$
{}~= ~\frac{1}{ \sqrt3}  A^{nf} (D^{+} \rightarrow \bar
 K^{0} \rho^{+} ). \eqno(23)$$
The last relation (23) leads to the following constraint:
$$ \frac {A_{1/2}^{nf}(D\rightarrow \bar K \rho) }
{A_{3/2}^{nf}(D\rightarrow\bar K \rho)}~~= ~~
\frac {c^{2}_{1} + 2c^{2}_{2}} {\sqrt2 (c^{2}_{2} -c^{2}_{1})}~
 = ~ -1.123 \pm 0.112.\eqno(24) $$
Experimental value $B(D^{+} \rightarrow \bar K^{0} \rho^{+})~ = ~
 6.6 \%$ then predicts sum of the branching ratios of
$D^{0} \rightarrow \bar K \rho$ modes:
$$ B(D^{0}\rightarrow K^{-}\rho^{+})~ +~
B(D^{0}\rightarrow \bar K^{0}\rho^{0}) ~=~
10.42 \%, ~~~ (11.50 \pm 1.31\%~ {\rm Expt.}) \eqno(25)$$
in good agreement with experiment. Present data [1] on these modes is
consistent with a choice of zero phase difference between isospin $1/2$ and
$3/2$
channels and fixes the ratio of nonfactorizable amplitudes,
$$\frac {A_{1/2}^{nf}(D\rightarrow \bar K \rho) }
{A_{3/2}^{nf}(D\rightarrow\bar K \rho)} ~ =~
 -1.481 \pm 0.582~(\rm {Expt.}) \eqno(26)$$
consistent with theoretical value given in Eq. (24). For branching ratio of
the individual modes, we obtain
$$ B(D^{0}\rightarrow K^{-}\rho^{+})
{}~ =~9.37\% ~~~(10.4 \pm 1.3\%~ {\rm Expt.}),
  \eqno(27)$$
$$ B(D^{0}\rightarrow \bar K^{0}\rho^{0}) ~ = ~
 1.05 \%,~~~ (1.10 \pm 0.18\%~ {\rm Expt.}). \eqno(28)$$
Performing a similar analysis for the $ D \rightarrow \bar K^{*} \pi$ decay
amplitudes and using the experiemental value $B(D^{+} \rightarrow
 \bar K^{*0} \pi^{+})~ = ~2.2 \%$, we calculate:
$$ B(D^{0}\rightarrow K^{*-}\pi^{+})~ + ~B(D^{0}\rightarrow \bar
K^{*0}\pi^{0})~ = ~7.44 \%,~~~ (7.9 \pm 0.7\%,~ {\rm Expt.})\eqno(29)$$
in nice agreement with experiment. For these modes also, the isospin reduced
amplitudes bear the same ratio as given in Eq.(24),
$$ \frac {A_{1/2}^{nf}(D\rightarrow \pi \bar K^{*}
) } {A_{3/2}^{nf}(D\rightarrow \pi \bar K^{*})} ~ = ~
\frac {c^{2}_{1} + 2c^{2}_{2}} {\sqrt2 (c^{2}_{2} -c^{2}_{1})} ~ =~
{}~ -1.123 \pm 0.112, \eqno(30) $$
which compares well with experimental value $ -1.171 \pm 0.158$ when negative
sign
is chosen for $A_{3/2}$.
\par
 Calculation of branching ratio of the remaining $D \rightarrow PV$
decays needs numerical values of the reduced amplitudes. Apparently these
decays seem to involve several unknown parameters. However,
the parameters $d$'s and $e$'s appear only when isosinglet meson is emitted.
Also that $D$ meson decays involve combinations $(a_{1}-b_{1})$ and
$(a_{2}-b_{2})$ which in fact are expressible as
$$ a_{1}-b_{1} ~=~ [-(c_{1}+c_{2})a_{3} + c_{2}b_{3} +
c_{1}b_{4}]/(c_{1}-c_{2}),$$
$$ a_{2}-b_{2} ~=~ [+(c_{1}+c_{2})a_{3} + c_{1}b_{3} +
c_{2}b_{4}]/(c_{1}-c_{2}), \eqno(31)$$
where
$$ a_{3} ~ = ~ -0.042 ~ GeV^{2},
\eqno(32)$$
$$ b_{3} + b_{4}~ =~ -0.251 ~ GeV^{2}, \eqno(33)$$
are given by $D^{+}$ modes. The relations (31) follow from the constraints
given in Eqs. (24) and (30). With the experimental
values $B(D^{0}\rightarrow \bar K^{0}\phi) ~=~ 0.83 \%, $  and
$ B(D^{0}\rightarrow \bar K^{0}\omega) ~ =~2.0 \%,$, and taking negative and
positive signs for their experimental amplitudes, we find (in $GeV^{2}$),
$$ b_{3} ~=~ 0.042, \eqno(34)$$
$$ b_{4} ~=~ -0.293,  \eqno(35)$$
$$ e_{1} + d_{1} ~ = ~ -0.047. \eqno(36)$$
The relations given in (31) then yield:
$$ a_{1} - b_{1} ~= ~ -0.202,\eqno(37)$$
$$ a_{2} - b_{2} ~=~ 0.108. \eqno(38)$$
Using the measured branching ratios $ B(D^{+}_{s}\rightarrow K^{+} \bar
K^{*0}) ~ = ~ 3.3 \% $ and
$ B(D^{+}_{s}\rightarrow \bar K^{0} K^{*+}) ~ = ~4.2 \%, $ we find (in
$GeV^{2}$),
$$ a_{1} + b_{1} ~ = ~0.190, \eqno(39)$$
$$ a_{2} + b_{2} ~ = ~0.123, \eqno(40) $$
by taking negative and positive signs of their experimental amplitudes
respectively. These parameters then predict
$$ B(D^{+}_{s}\rightarrow \pi^{0} \rho^{+}) ~=~ B(D^{+}_{s}\rightarrow
 \pi^{+} \rho^{0}) ~ =~ 0.10 \% \eqno(41)$$
which is well below the experimental upper limit ($ < ~0.28 \%$) for
$B(D^{+}_{s}\rightarrow \pi^{+} \rho^{0})$.
Further, experimental value $B(D\rightarrow \pi^{+}
\phi)  ~ = ~ 3.5 \% $ yields,
$$ e_{1} - d_{1} ~=~ 0.070 ~GeV^{2}, \eqno(42) $$
for the positive choice of its experimental
amplitude. This value in turn leads to: $$ B(D\rightarrow
\pi^{+} \omega) ~ =~ 0.61 \% \eqno(43)$$
obeying the exprimental upper limit $(<~1.7 \%)$.
Now, we are left with $\eta$ and $\eta'$ emitting decays which invlove
mixing angle $\phi_{Phy}$. We have chosen to present results for all the three
mixing angles
$-10^{0}, ~-19^{0},$ and $-23^{0}$ given in the Particle Data Group [1] so as
to
make the trend with mixing angle evident. Measured
branching ratios $ B(D^{0} \rightarrow \eta \bar K^{*0}) ~= ~ 1.9 \% $
and $ B(D_{s}^{+} \rightarrow \eta'
\rho^{+}) ~ =~ 12.0 \% $ fix the parameters:
$$ e_{2} + d_{2} ~ = ~0.668~ GeV^{2} ~~ {\rm for} ~ \phi_{Phy} ~ =~ -10^{0} $$
$$ \hskip 2cm = ~ 0.373 ~GeV^{2} ~~ {\rm for} ~ \phi_{Phy} ~ = ~-19^{0},$$
$$ \hskip 2cm = ~0.313 ~ GeV^{2}~~ {\rm for} ~\phi_{Phy} ~= ~ -23^{0},
\eqno(44)$$
and
$$ e_{2} - d_{2} ~ = ~0.457~GeV^{2}~~ {\rm for } ~\phi_{Phy}~= ~ -10^{0} $$
$$\hskip 2cm = ~0.428~ GeV^{2} ~~{\rm for}~\phi_{Phy} ~=~-19^{0},$$
$$\hskip 2cm = ~0.412~GeV^{2}~~ {\rm for} ~ \phi_{Phy}~= ~-23^{0}, \eqno(45)$$
for negative and positive signs of respective experimental amplitudes.
Calculated branching ratios for $\eta$ and $\eta'$ emitting
decays are listed in columns (ii) to (iv) of Table III for different values of
the mixing angles. For the sake of comparison with factorizable part,
nonfactorizable
contributions to various decays are given in column (iii) of Table II.
\vskip 1.0cm
\par
\noindent {\bf Acknowledgments}
\par
\noindent
One of the authors (RCV) is grateful to Professor A. Salam, the President, and
Professor
G. Virasoro, the Director, ICTP, Trieste, UNESCO, and Dr. S. Randjbar-Daemi for
the
hospitality at the ICTP, Trieste.
\newpage
\newpage
\begin{center}
Table I \\
Spectator-quark decay amplitudes
\end{center}
\begin{center} \baselineskip 24pt
\large \begin{tabular}{| c| c| c|}
\hline
Process & Amplitude $(A^{f})$ & $a_{1}=1.09,~a_{2}=-0.09$  \\
\hline
& & \\
$ D^{0} \rightarrow K^{-} \rho^{+} $ & $ 2a_{1}f_{ \rho}
m_{\rho}F_{1}^{DK}(m_{ \rho}^{2})$  & 0.316 \\
$ D^{0} \rightarrow \bar
K^{0} \rho^{0}$ & $ \sqrt2 a_{2}f_{K}
m_{\rho}A_{0}^{D\rho}(m_{K}^{2})$ & -0.011  \\
$ D^{0} \rightarrow \bar
K^{0} \omega $ & $ \sqrt2 a_{2}f_{K} cos \theta'
m_{\omega}A_{0}^{D
\omega}(m_{K}^{2})$  & -0.011 \\
$ D^{0} \rightarrow \bar K^{0} \phi$ &
$ - \sqrt2 a_{2} sin \theta' f_{K} m_{\phi}A_{0}^{D
\phi}(m_{K}^{2})$ & 0 \\
$ D^{0} \rightarrow \pi^{+} K^{*-}$ &  $ 2a_{1} f_{\pi}
m_{K^{*}}A_{0}^{DK^{*}}(m_{\pi}^{2})$ & 0.179\\
$ D^{0} \rightarrow
\pi^{0} \bar K^{*0} $ &  $ \sqrt2 a_{2} f_{K^{*}} m_{K^{*}}F_{1}^{D\pi}
(m_{K^{*}}^{2})$ & -0.026\\
$ D^{0} \rightarrow \eta \bar K^{*0}$ & $
\sqrt2 a_{2} f_{K^{*}} sin \theta m_{K^{*}} F_{1}^{D\eta} (m_{K^{*}}^{2})$
& -0.015\\
$ D^{0} \rightarrow \eta' \bar K^{*0}$ & $ \sqrt2 a_{2} f_{K^{*}} cos
\theta m_{K^{*}} F_{1}^{D\eta'} (m_{K^{*}}^{2})$ &-0.015 \\
 & & \\
$ D^{+}\rightarrow \bar
K^{0} \rho^{+} $ & $ 2a_{1} f_{\rho}
m_{\rho} F_{1}^{DK}(m_{\rho}^{2}) $ &   \\
 & $ + 2a_{2} f_{K}m_{\rho}A_{0}^{D\rho}(m_{K}^{2})$ & 0.300 \\
$D^{+}\rightarrow
\pi^{+} \bar K^{*0} $ & $ 2a_{1} f_{\pi}
m_{K^{*}}A_{0}^{DK^{*}}(m_{\pi}^{2}) $ & \\
 & $ + 2a_{2} f_{K^{*}}m_{K^{*}}F_{1}^{D\pi}(m_{K^{*}}^{2})$ & 0.143 \\
 & & \\
$ D^{+}_{s}\rightarrow
\pi^{+} \rho^{0} $ & $ 0 $ & 0  \\
$ D^{+}_{s} \rightarrow \pi^{0}
\rho^{+} $ & $ 0 $  & 0 \\
$D^{+}_{s} \rightarrow \pi^{+} \omega $ & $2a_{1} f_{\pi} sin
\theta'm_{\omega}A_{0}^{D_{s}\omega}(m_{\pi}^{2})$ & 0 \\
$D^{+}_{s} \rightarrow \pi^{+} \phi $ & $ 2a_{1} f_{\pi} cos
\theta'm_{\phi}A_{0}^{D_{s}\phi}(m_{\pi}^{2})$   & 0.204 \\
$ D^{+}_{s}\rightarrow \eta \rho^{+} $ & $ -2a_{1} f_{\rho} cos \theta
m_{\rho}F_{1}^{D_{s}\eta}(m_{\rho}^{2})$ & -0.209 \\
$ D^{+}_{s}
\rightarrow \eta' \rho^{+} $ & $ 2a_{1} f_{\rho} sin \theta
m_{\rho}F_{1}^{D_{s}\eta'}(m_{\rho}^{2})$ & 0.205  \\
$D^{+}_{s}
\rightarrow K^{+} \bar K^{*0} $ & $ 2a_{2} f_{K^{*}}
m_{K^{*}}F_{1}^{D_{s}K}(m_{K^{*}}^{2})$ & -0.033 \\
$ D^{+}_{s} \rightarrow \bar K^{0} K^{*+} $ & $ 2a_{2} f_{K}
m_{K^{*}}A_{0}^{D_{s}K^{*}}(m_{K}^{2})$ & -0.018 \\
  & & \\
\hline
\end{tabular}
\end{center}
\newpage
\begin{center}
Table II \\
Nonfactorizable contributions to $ D \rightarrow PV $ decays.\\
($ c_{1}$ and $c_{2}$ are the QCD coefficients)
\end{center}
\begin{center}
\normalsize \baselineskip 24pt \begin{tabular}{ | c|c| c|}
\hline Process   &  Nonfactorizable contribution  & $\phi=-10^{0}$ \\
\hline
& & \\
$D^{0} \rightarrow K^{-} \rho^{+} $ & $ c_{2} [-a_{1} - a_{3} +
b_{1} + b_{4}]$ & 0.024 \\
$ D^{0} \rightarrow \bar K^{0} \rho^{0}$
& $ \frac {1} {\sqrt2}c_{1} [a_{1} - a_{3} - b_{1} + b_{3}]$  & -0.103 \\
$ D^{0} \rightarrow \bar K^{0} \omega $ & $ c_{1}[\frac {cos \theta'} {\sqrt2}
(-a_{1} - a_{3} + b_{1} + b_{3} + 2(e_{1} + d_{1})) + sin
\theta'
(-a_{2} + b_{2} + e_{1} + d_{1})]$ & 0.174 \\
$ D^{0} \rightarrow \bar
K^{0} \phi$ & $ c_{1}[\frac {sin \theta'} {\sqrt2} (-a_{1} -
a_{3} + b_{1} + b_{3} + 2(e_{1} + d_{1})) - cos \theta' (a_{2} -
b_{2} - e_{1} - d_{1})]$  & -0.199 \\
$ D^{0} \rightarrow \pi^{+}
K^{*-}$ & $ c_{2}[-a_{2} + a_{3} + b_{2} + b_{3}]$ & 0.058 \\
$D^{0} \rightarrow
\pi^{0} \bar K^{*0} $ &  $ \frac {1} {\sqrt2} c_{1} [a_{2} + a_{3} -
b_{2} + b_{4}]$  & -0.201 \\
$ D^{0} \rightarrow \eta \bar K^{*0}$ &
$ c_{1}[\frac {sin \theta} {\sqrt2} (-a_{2} + a_{3} + b_{2} +
b_{4} + 2(e_{2} + d_{2})) + cos \theta (a_{1} - b_{1} - e_{2} -
d_{2})]$  & -0.210 \\
$ D^{0} \rightarrow \eta' \bar K^{*0}$ & $ c_{1}[\frac {cos \theta}
{\sqrt2} (-a_{2} + a_{3} + b_{2} + b_{4} + 2(e_{2} + d_{2})) +
sin \theta (-a_{1} + b_{1} + e_{2} + d_{2})]$ & 1.334  \\
 & & \\
$ D^{+}\rightarrow
\bar K^{0} \rho^{+} $ & $ (c_{1} + c_{2})[-2a_{3} + b_{3} + b_{4}]$  &-0.122 \\
$ D^{+}\rightarrow \pi^{+} \bar K^{*0} $ & $ (c_{1} + c_{2})[2a_{3} +
b_{3} +b_{4}]$ & -0.254\\
 & & \\
$ D^{+}_{s} \rightarrow \pi^{+} \rho^{0} $ & $
\frac {1} {\sqrt2} c_{2} [-a_{1} + a_{2} - b_{1} + b_{2}]$  & 0.024 \\
$ D^{+}_{s} \rightarrow \pi^{0} \rho^{+} $ & $ \frac {1} {\sqrt2}
c_{2} [a_{1} - a_{2} + b_{1} - b_{2}]$ & 0.024  \\
$ D^{+}_{s} \rightarrow
\pi^{+} \omega $ & $  c_{2}[\frac {cos \theta'} {\sqrt2} (a_{1} + a_{2}
+ b_{1} + b_{2} - 2(e_{1} - d_{1})) + sin \theta' (a_{3} + b_{3}
- e_{1} + d_{1})]$  & -0.062 \\
$ D^{+}_{s} \rightarrow \pi^{+} \phi
$ & $ c_{2}[\frac {sin \theta'} {\sqrt2} (a_{1} + a_{2} + b_{1}
+ b_{2} - 2(e_{1} - d_{1})) - cos \theta' (-a_{3} - b_{3} +
e_{1} - d_{1})]$  & 0.037 \\
$ D^{+}_{s} \rightarrow \eta \rho^{+} $ & $  c_{2}[\frac {sin
\theta} {\sqrt2} (a_{1} + a_{2} + b_{1} + b_{2} - 2(e_{2} - d_{2})) +
cos \theta (a_{3} - b_{4} + e_{2} - d_{2})]$  & -0.097 \\
$ D^{+}_{s}
\rightarrow \eta' \rho^{+} $ & $  c_{2}[\frac {cos \theta} {\sqrt2}
(a_{1} + a_{2} + b_{1} + b_{2} - 2(e_{2} - d_{2})) + sin \theta
(-a_{3} + b_{4} - e_{2} + d_{2})]$ & 0.424 \\
$ D^{+}_{s} \rightarrow K^{+}
\bar K^{*0} $ & $ c_{1}[a_{1} + a_{3} + b_{1} + b_{4}]$   & -0.186 \\
$ D^{+}_{s} \rightarrow \bar K^{0} K^{*+} $ & $ c_{1}[a_{2} -
a_{3} + b_{2} + b_{3}]$ & 0.263 \\
 &  & \\
\hline
\end{tabular}\end{center}
\newpage \large
\begin{center}
Table III \\
Branching ratios (\%) of $ \eta/ \eta' $ emitting decays for
different mixing angles.\\
\end{center}
\begin{center} \normalsize
\begin{tabular}{ | c| c | c | c | c | } \hline Decay & $
\phi_{phy} = -10^{o}$ &  $ \phi_{phy} = -19^{o} $ & $
\phi_{phy} = -23^{o} $  & Experiment \\
\hline & & & & \\
$ D^{0} \rightarrow \eta \bar K^{*0} $ & 1.90$^{a}$ & 1.90$^{a}$ & 1.90$^{a}$
& 1.9
$ \pm$ 0.4 \\
 & & & &  \\
$ D^{0} \rightarrow  \eta' \bar K^{*0}$ & 0.33 & 0.10 & 0.08 & $ < 0.11 $ \\
 & & & &  \\
$ D_{s}^{+} \rightarrow \eta \rho^{+} $
 & 10.47  & 5.23 & 3.50 & 10.0 $ \pm$ 2.2 \\
 & & & & \\
 $ D_{s}^{+}\rightarrow \eta' \rho^{+}$
&  12.00$^{a}$ &12.00$^{a}$ &12.00$^{a}$ & 12.0 $ \pm$ 3.0 \\
 & & & & \\
\hline
\end{tabular}
\end{center}
$^{a}$ input
\newpage \large


\newpage \baselineskip24pt
\begin {thebibliography} {99}
\bibitem[1]{} L. Montanet et al., Particle data group, Phys. Rev. D {\bf 50},
3-I (1994).
\bibitem[2] {} M. Bauer, B.Stech and
M. Wirbel, Z. Phys. C {\bf 34}, 103 (1987); M. Wirbel, B. Stech and M. Bauer,
Z. Phys. C {\bf 29}, 637 (1985).
\bibitem[3] {} N. Isgur, D.
Scora, B. Grinstein and M. Wise, Phys. Rev. D {\bf 39}, 799 (1989).
\bibitem [4] {} S. Stone, in `Heavy flavors', A.J. Buras and M. Linder (eds),
 World Sci. Pub. Singapore (1992).
\bibitem[5] {} M. Gourdin, A. N. Kamal, Y. Y. Keum and X. Y. Pham, Phys.
Letts. B {\bf 333}, 507 (1994); A. N. Kamal and T. N. Pham, Phys. Rev. D {\bf
50},
395 (1994).
\bibitem[6] {} CLEO collaboration: M.S. Alam {\it et al.}, Phys. Rev.
D {\bf 50}, 43 (1994); D. G.  Cassel, `Physics from CLEO', talk
delivered at Lake-Louise Winter Institute on `Quarks and Colliders',
Feb. (1995).
\bibitem[7] {} A. N. Kamal, Q. P. Xu, and A. Czarnecki, Phys. Rev. D
{\bf 49}, 1330 (1990); R. C. Verma, A. N. Kamal and M. P. Khanna,
Z. Phys. C. {\bf 65}, 255 (1995).
\bibitem[8]{} R. C. Verma, `A Puzzle in $
D, D_{s} \rightarrow \eta / \eta' + P/V'$, talk delivered at Lake Louise
Winter Institute on `Quarks and Colliders' Feb. (1995).
\bibitem[9] {} N. G. Deshpande, M. Gronau, and D. Sutherland, Phys. Letts.
 {\bf 90 B}, 431 (1980).
\bibitem[10] {} H. Y. Cheng, Z. Phys. C. {\bf 32}, 237 (1986), `Nonfactorizable
contributions to nonleptonic Weak Decays of Heavy Mesons', IP-ASTP-
\underline {11} -94, June (1994); J. M. Soares, Phys. Rev. D {\bf 51}, 3518
(1995).
\bibitem[11] {} A. N. Kamal and A. B. Santra, `Nonfactorization and color
Suppressed $ B
\rightarrow \psi ( \psi (2S)) + K ( K^{*})$ Decays, University of
Alberta preprint (1995); Nonfactorization and the Decays $ D_{s}^{+}
\rightarrow \phi \pi^{+}, \phi \rho^{+},$ and $ \phi e^{+} \nu_{e}$
Alberta-Thy-1-95, Jan (1995); A. N. Kamal, A. B. Santra, T. Uppal and R.
C. Verma, `Nonfactorization in Hadronic two-body Cabibbo favored decays
of $ D^{0}$ and $ D^{+}$, Alberta-Thy-08-95, Feb. (1995).
\bibitem[12] {} R. C. Verma, Zeits. Phys. C (1995) {\it in press}
\bibitem[13] {} R. C. Verma, `SU(3)-flavor analysis of nonfactorizale
contributions to $D \rightarrow PP$ decays', Panjab Univ.-April 95.
\bibitem[14]{} R. C. Verma and A. N. Kamal, Phys. Rev. D, {\bf 43}, 829 (1990).
\bibitem[15]{} L.L Chau and H. Y. Cheng, Phys. Lett. {\bf B 333}, 514 (1994).
\bibitem[16]{} M. S. Witherall, International Symposium on Lepton and
Photon Interactions at High Energies, Ithaca, N.Y. (1993), edited by
P. Drell and D. Rubin, AIP Conf. Proc. No. 302 (AIP, New York) p. 198.
\bibitem[17]{} A. N. Kamal and T. N. Pham, Phys. Rev. D, {\bf 50}, 6849 (1994);
ibid {\bf 50}, R1832 (1994).
\end {thebibliography}
\end {document}